\documentclass{llncs}

\usepackage{makeidx} 
\usepackage{graphicx}

\begin{document}

\frontmatter          % for the preliminaries
\pagestyle{headings} 

\title{Blind Signal Separation Methods for the Identification of Interstellar Carbonaceous Nanoparticles
\thanks{This work is based on observations made with the Spitzer Space
Telescope, which is operated by the Jet Propulsion Laboratory, California Institute of Technology 
under a contract with NASA.}}

\author{O. Bern\'e \inst{1,2}
	\and
	Y. Deville\inst{2}
	\and 
	C. Joblin\inst{1}}

\institute{
Centre d'Etude Spatiale des Rayonnements, CNRS et Universit\'e Paul 
Sabatier Toulouse~3, Observatoire Midi-Pyr\'en\'ees, 9 Av. du Colonel Roche, 
31028 Toulouse cedex 04, France,\\
\email{olivier.berne@cesr.fr}
,
\email{christine.joblin@cesr.fr}
\and
Laboratoire d'Astrophysique de Toulouse-Tarbes, CNRS et Universit\'e Paul 
Sabatier Toulouse~3, Observatoire Midi-Pyr\'en\'ees, 14 Av. Edouard Belin,
31400 Toulouse, France,\\
\email{ydeville@ast.obs-mip.fr}}

\maketitle

\begin{abstract}
The use of Blind Signal Separation methods (ICA and other approaches)
for the analysis of astrophysical data remains quite unexplored. In this paper, 
we present a new approach for analyzing
%YDsupprime car redite: the analysis of 
the infrared emission spectra of interstellar
dust, obtained with NASA's Spitzer Space Telescope, using \emph{FastICA} and 
Non-negative Matrix Factorization (NMF). 
Using these two methods, we were able to unveil 
the \emph{source} spectra of three different types of carbonaceous nanoparticles
present in interstellar space.
These spectra can then constitute a basis for the interpretation of the mid-infrared 
emission spectra of interstellar dust in the Milky Way and nearby galaxies.
We also show how to use these extracted spectra to derive the spatial
distribution of these nanoparticles.
\end{abstract}

%%%%%%%%%%%%%%%%%%%%%%%%%%%%%%%%%%%%%%%%%%%
\section{Introduction} \label{introduction}
%%%%%%%%%%%%%%%%%%%%%%%%%%%%%%%%%%%%%%%%%%%

The Spitzer Space Telescope (\emph{Spitzer})
%YDsupprime: , see Werner 2004) 
comprises one of today's best 
instruments to probe the mid-infrared (mid-IR) emission of interstellar dust
in the Milky Way and nearby galaxies.
This emission is mainly carried by very small (nanometric) interstellar dust particles. 
One of the goals of infrared astronomy is to identify the physical/chemical nature 
of these species, as they play a fundamental role in the evolution of galaxies. 
Unfortunately, the observed spectra are mixtures of the emission  from various dust populations.
The strategy presented in this paper is to apply Blind Signal Separation (BSS) methods i.e. \emph{FastICA}
and NMF to a set of \emph{Spitzer} mid-IR (5-30~$\mu$m) spectra obtained with the 
InfraRed Spectrograph (IRS),
in order to extract the genuine spectrum of each population of nanoparticles. 
We first present these observations in Sect. \ref{observations}, then we
apply the two BSS methods to these observations and finally give an example of
how the extracted spectra can be used to trace the evolution of dust, in the
Milky Way and external galaxies.

%%%%%%%%%%%%%%%%%%%%%%%%%%%%%%%%%%%%%%%%%%%
\section{Observations} \label{observations}
%%%%%%%%%%%%%%%%%%%%%%%%%%%%%%%%%%%%%%%%%%%

\begin{figure}[!h]
\begin{center}
\includegraphics[width=11cm]{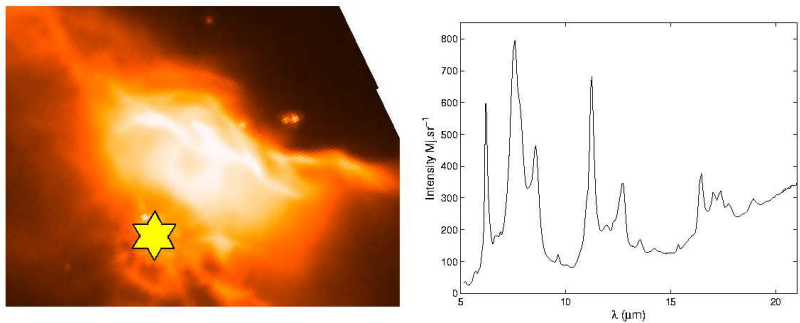}
\vspace{-0.5cm}
\caption{\emph{Left}: Infrared (8~$\mu$m) view of the NGC 7023 North PDR. 
The star is illuminating the cloud situated in the upper part of the image. 
\emph{Right}: Mid-IR spectrum for a given
position in the spectral cube of NGC 7023. 
}
\label{pdr}
\vspace{-0.5cm}
\end{center}
\end{figure}

We have observed with \emph{Spitzer} nearby photo-dissociation regions
(PDRs), which consist of a star illuminating the border of dense clouds of gas and dust.
The physical conditions (UV field intensity and hardness, cloud density) strongly vary from a PDR
to another as well as inside each PDR depending on the considered position. These variations
are extremely useful to probe the nature of dust particles which are altered
by the local physical conditions \cite{rap05}.
Therefore, we have observed 
%YDsupprime: a sample of 
11 PDRs as part of the SPECPDR program
using the IRS in "spectral mapping" mode. This mode enabled us to construct one dataset for each
PDR. This dataset is a spectral cube, with two spatial dimensions and one spectral dimension (see 
Fig.~\ref{pdr}). Each spectral cube is thus a 3-dimensional matrix $C (p_x, p_y, \lambda )$, 
which defines the variations of the recorded data with respect to the wavelength $\lambda $, 
for each considered position with coordinates $(p_x, p_y)$ in the cube. The dimensions of these
cubes are generally about $30\times 30$ positions and $250$ points in wavelength
ranging between 5 and 30 $\mu$m.

%%%%%%%%%%%%%%%%%%%%%%%%%%%%%%%%%%%%%%%%%%%%%%%%%%%%%%%%%%%%%%%%%%%%%%%%%%
\section{Blind Separation of Interstellar Dust Spectra} \label{bss}
%%%%%%%%%%%%%%%%%%%%%%%%%%%%%%%%%%%%%%%%%%%%%%%%%%%%%%%%%%%%%%%%%%%%%%%%%%

BSS is commonly used to restore a set of 
unknown "source" signals from a set of observed signals which are mixtures of these 
source signals, with unknown mixture parameters \cite{hyv01}. BSS is most often 
achieved using ICA methods such as \emph{FastICA} \cite{hyv99}. An alternative class of 
methods for achieving BSS is NMF, which was introduced in \cite{lee99} and then 
extended by a few authors. In the astrophysical community, ICA has been successfully 
used for spectra discrimination in infrared spectro-imagery of Mars ices \cite{for05}, 
elimination of artifacts in astronomical images \cite{fun03} or extraction of cosmic 
microwave background signal in \emph{Planck} simulated data \cite{mai02}. To our 
knowledge, NMF has not yet been applied to astrophysical problems. However, it has 
been used to separate spectra in other application fields, e.g. for magnetic resonance 
chemical shift imaging of the human brain \cite{saj04} or for analyzing wheat grain 
spectra \cite{gob05}.

The simplest version of the BSS problem concerns so-called "linear instantaneous" 
mixtures. It is modeled as follows:
\begin{equation}
\label{general}
X=AS
\end{equation}
where $X$ is an $m\times n$ matrix containing $n$ samples of $m$ observed signals, $A$ is 
an $m\times r$ mixing matrix and $S$ is an $r \times n$ matrix containing $n$ samples of 
$r$ source signals. The observed signal samples are considered to be linear combinations 
of the source signal samples (with the same sample index). It is assumed that $r \leq m$
in most investigations, including this paper. The objective of BSS algorithms is then to 
recover the source matrix $S$ and/or the mixing matrix $A$ from $X$,
up to BSS indeterminacies.

The correspondence between the generic BSS data model (\ref{general}) and the 3-dimensional 
spectral cube $C (p_x, p_y, \lambda )$ to be analyzed in the present paper may be defined 
as follows. In this paper, the sample index is associated to the wavelength $\lambda$, 
and each observed signal consists of the overall spectrum recorded for a cube pixel $(p_x, p_y)$.
Each one of these signals defines a row of the matrix $X$ in Eq. (\ref{general}).
Moreover, each observed spectrum is a linear combination of "source spectra" 
(see Sect.~\ref{suit}), which are respectively associated to each of the (unknown) 
types of nanoparticles that contribute to the recorded spectral cube. Therefore, 
the recorded spectra may here be expressed according to (\ref{general}), with 
unknown combination coefficients in $A$, unknown source spectra in $S$ and 
an unknown number $r$ of source spectra.

\subsection{Suitability of BSS Methods for the Analysis of \emph{Spitzer}-IRS Cubes}
\label{suit}

In order to apply the NMF or \emph{FastICA} to the IRS data cubes, it is necessary to 
make sure that the "linear instantaneous" mixture condition is fulfilled. 
Here we consider that each observed spectrum is a linear
combination of "source spectra", which are due to the emission of different populations of dust nanoparticles. The main effect that can disturb the linearity of the model is radiative
 transfer as shown by \cite{nuz}, because of the non-linearity of the equations. In our 
case however, this effect is completely negligible because the emission
spectra we observe come from the surface of clouds and are therefore not altered by 
radiative transfer. 

\subsection{Considered BSS Methods}
%YDsupprime: Application to \emph{Spitzer} data}
\label{appli}

In this section, we detail which particular BSS methods we have applied to the observed data.
%YDsupprime: , and how we proceeded.

\subsubsection{NMF}
\label{nmfappli}

We used NMF as presented in \cite{lee01}. The matrix of observed spectra $X$ is 
approximated using

\begin{equation}
W
%YDsupprime: \times 
H,
\end{equation}
where $W$ and $H$ are non-negative
matrices, with the same dimensions as in (\ref{general}). 
This approximation is optimized by adapting the matrices 
$W$ and $H$ using the algorithm of  \cite{lee01} in order to minimize the divergence 
between $X$ and 
$W
%YDsupprime: \times 
H$. We implemented the algorithm with Matlab.
Convergence is 
reached after about 1000 iterations (which takes less than one minute with a 3.2 GHz 
processor). The value of $r$ (number of "source" spectra) is not imposed by the NMF method.
Our strategy for setting it so as to extract the sources was the following:
\vspace{0.5cm}

$\bullet$ Apply the algorithm to a 
given dataset, with the minimum number of assumed sources,
i.e. $ \hat{r} = 2$, providing 2 sources.

$\rightarrow$ If the found solutions are physically coherent and
linearly independent, we consider that at least $\hat{r}=2$ sources can be extracted.

$\rightarrow$ Else, we consider that the algorithm is not suited for analysis
(this never occurred in our tests).

\vspace{0.5cm}
$\bullet$ Try the algorithm on the same dataset but with $\hat{r}=3$ sources.

$\rightarrow$ If the found solutions are physically coherent and
linearly independent, we consider that at least $\hat{r}=3$ sources can be extracted.

$\rightarrow$ Else, we consider that
only two sources can be extracted, extraction was over with $\hat{r}=2$ and thus $r=2$.

\vspace{0.5cm}
$\bullet$ Same as previous step but with $\hat{r}=4$ sources.

$\rightarrow$ If the found solutions are physically coherent and
 linearly independent, we consider that at least $\hat{r}=4$ sources can be extracted.

$\rightarrow$ Else we consider that 
only three sources can be extracted, extraction was over with $\hat{r}=3$ and thus $r=3$.

$\ldots$

Physically incoherent spectra exhibit sparse peaks (spikes) 
which cannot be PDR gas lines. We found $r=3$ for NGC 7023-NW and $r=2$ for the other PDRs, 
implying that we could respectively extract 3 and 2 spectra from these data cubes.

\subsubsection{\emph{FastICA}}
\label{fasticaappli}

We used \emph{FastICA} in the deflation version \cite{hyv99}
in which each source is 
extracted one  after the other and subtracted from the observations until all 
sources are extracted. The advantage of this
\emph{FastICA} method is that it is not necessary 
to fix, before running the algorithm,
the number $r$ of sources that we want to extract, as it is for NMF. 
The extraction of the sources takes less than one minute using \emph{FastICA} 
coded with Matlab, and with a 3.2 GHz processor.

\subsection{Results}
\label{res}

\begin{figure}[!h]
\includegraphics[width=\hsize]{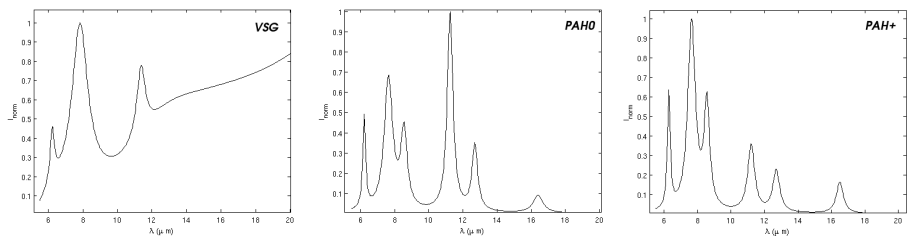}
\vspace{-0.5cm}
\caption{ The three BSS-extracted spectra from our study on PDRs.}
\label{spec}
\vspace{-0.5cm}
\end{figure}

Using the BSS methods presented in this paper, we were able to extract up to three source 
spectra from the \emph{Spitzer} observations. The number $r$ of sources found in a given 
PDR is always the same with NMF and \emph{FastICA}.
The three extracted spectra in NGC 7023 North are presented in Fig.~\ref{spec}. 
Two of them exhibit the series of aromatic bands which have previously been attributed 
to Polycyclic Aromatic Hydrocarbons (PAHs, \cite{leg84} and \cite{all85}).
These two spectra show different band intensity ratios. One is the spectrum of neutral PAHs
(PAH$^0$) while the other is due to ionized PAHs (PAH$^+$).
The last spectrum exhibits a continuum and aromatic bands, which can be attributed to
very small carbonaceous grains (VSGs), possibly PAH clusters \cite{rap06}.

\subsection{\emph{FastICA} vs NMF for our Application}
\label{vs}

As mentioned in 
%YDsupprime: Sect.~\ref{appli}, 
Sect.~\ref{res},
we were able to extract the source spectra from our 
data using both \emph{FastICA} and NMF. 
%YDsupprime: However the extractions are not exactly similar. 
However, the extracted spectra are not exactly the same for both
methods.
We conducted several tests in order to be able to evaluate which one of the two 
methods is more appropriate for our application. 
We created a set of 2/3 artificial 
carbonaceous nanoparticle spectra, to which we added a variable level of white, spatially 
homogeneous noise. We mixed these spectra with a random matrix to create a set of 100 
artificial observed spectra. We then applied the two BSS methods considered in this paper. 
With a noise level at zero, both methods recover the original signals
with high efficiency (correlation coefficients between original and extracted
%YDsupprime: unmixed 
signals above 0.995). When 
adding noise, this efficiency decreases but remains acceptable down to a noise level
corresponding to a SNR of 3dB (which is much lower that the average SNR of the \emph
{Spitzer spectra}). We note however that the efficiency of FastICA drops slightly faster 
than the one of NMF under the effect of an increasing noise, and drops dramatically 
below a SNR of 3dB, while NMF can still partly recover the original signals. Finally,
with both methods we observe that the power of the residuals (i.e. observed signal minus
signal reconstructed from the estimated sources and mixing coefficients) has the same level 
as Spitzer noise.

We have shown in Sect.~\ref{res} that there are two main populations: one with a continuum (VSGs) 
and one with bands only (PAHs). Using \emph{FastICA}, we sometimes find a residual continuum in 
the BSS-extracted PAH spectrum, which we interpret as an incomplete separation.
It is possible that the criterion of NMF is more appropriate in our case because less 
restrictive. Indeed, NMF only requires non-negativity of the 
%YDsupprime: our observations
sources
and mixing coefficients, which is in 
essence the case for emission spectra, while \emph{FastICA} is based on the statistical independence and non-gaussianity 
of the sources, which is more difficult to prove. 
As a conclusion, we would like to stress the fact that both methods are very efficient for 
the first
task presented in this paper. We however note that NMF seems slightly better for this 
particular application.

%%%%%%%%%%%%%%%%%%%%%%%%%%%%%%%
\section{Deriving the Spatial Distribution of Carbonaceous Nanoparticles}
%%%%%%%%%%%%%%%%%%%%%%%%%%%%%%%

The next step of our analysis consists in using our extracted source spectra (Fig.~\ref
{spec}) in order to determine the spatial distribution of the three populations in galactic 
clouds or in external galaxies. The \emph{Spitzer} observations on-line archive contains 
hundreds of mid-IR spectral cubes of such regions which can be interpreted in this way. Our 
strategy consists in calculating the correlation parameter
$
%YDpasmiscarmanqueplace: \begin{equation}
c_p=E[Obs(p_x, p_y, \lambda) y_p ( \lambda) ]
%YDpasmiscarmanqueplace: \end{equation}
$
between an observed spectrum $Obs(p_x, p_y, \lambda)$ at a position $(p_x, p_y)$  in a spectral cube and one of our extracted source 
spectra $y_p (\lambda) $,
where $E[.]$ stands for expectation.
With the considered 
(i.e. linear instantaneous) mixture model, each observed spectrum reads
\begin{equation}
\label{specr}
Obs(p_x, p_y, \lambda)= \sum_{n} w(p_x, p_y)_{n} {S_n(\lambda)} 
\end{equation}
where $S_n(\lambda)$ is the 
$n^{th}$
source spectrum 
and $w(p_x, p_y)_{n}$ are the mixing
coefficients associated to that source.
Moreover,
BSS methods 
extract the sources up to arbirary scale factors, i.e. they
provide $y_p(\lambda)=\eta_p S_p(\lambda)$, where $\eta_p$ is 
an unknown scale
factor and $S_p(\lambda)$ is the 
$p^{th}$
source.
By
centering the observations 
and thus the extracted spectra,
and assuming that the sources are not correlated,
%YDpasmis: (\ref{specr}) shows that
the above-defined correlation 
parameter
becomes
\begin{equation}
\label{corr}
c_p= \eta_p w(p_x, p_y)_{p} E[S_p(\lambda)^2] 
\ .
\end{equation}

\begin{figure}[t]
\begin{center}
\includegraphics[width=8cm]{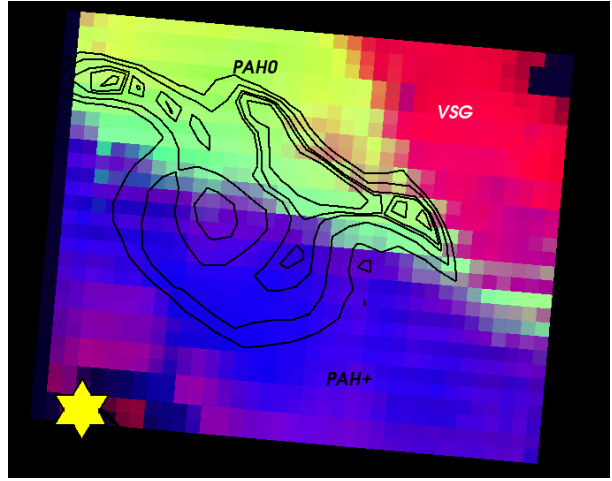}
\vspace{-0.0cm}
\caption{ Correlation maps of the three populations of nanoparticles in NGC 7023 North: VSGs in red,
PAH$^0$ in green and PAH$^+$ in blue. The contours in black
show the emission at 8 $\mu$m from Fig.~\ref{pdr}. The slight correlation of VSGs
with observations seen near the star is an artifact.}
\label{map}
\vspace{-0.3cm}
\end{center}
\end{figure}

\begin{figure}[!h]
\begin{center}
\includegraphics[width=10cm]{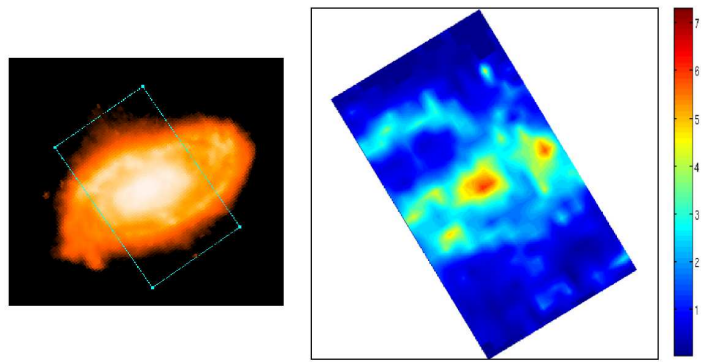}
\vspace{-0.3cm}
\caption{\emph{Left}: Infrared (8 $\mu$m) view of the NGC 4826 (Evil Eye) Galaxy. 
The rectangle indicates the region observed in spectral mapping with IRS. \emph{Right}:
Map of the ratio of $PAH^0$ over $PAH^+$ in NGC 4826 achieved using the
BSS-extracted spectra (Fig.\ref{spec}).}
\label{galaxy}
\vspace{-0.3cm}
\end{center}
\end{figure}

This coefficient $c_p$ is calculated for all the positions $(p_x, p_y)$,
therefore 
yielding a 2D correlation map.
Eq (\ref{corr}) shows that this map is proportional to
$w(p_x, p_y)_{p}$ and thus defines the spatial distribution of
the considered extracted source
$y_p(\lambda)=\eta_p S_p(\lambda)$.
We applied this approach to 
%YDsupprime: We did this first using 
the spectral cube of NGC 7023 North (Fig.~\ref{pdr}) and obtained 
the correlation maps presented in Fig.~\ref{map}. We find that the three nanoparticle 
populations emit in very different regions. It appears from the maps of Fig~\ref{map} 
that there is an evolution from a population of VSGs to PAH$^0$ and then $PAH^+$ while 
approaching the star. This reveals the processing of the nanoparticles by the UV 
stellar radiation.
The same strategy was tested using the cubes of external galaxies from the \emph{SINGS} program
which provides a database of mid-IR spectral cubes for
tens of nearby galaxies.
Fig.~\ref{galaxy} presents a map of the ratio of the two
correlation parameters, resp. of PAH$^0$ 
and PAH$^+$, obtained for the Evil Eye galaxy. This method provides a unique way 
to spatially trace the ionization fraction of PAHs which, combined with other tracers, is 
fundamental to understand the evolution of galaxies.

%YDsupprime: \clearpage

%%%%%%%%%%%%%%%%%%%%%%%%%%%%%%%%%%%%%%%
\section{Conclusion}
%%%%%%%%%%%%%%%%%%%%%%%%%%%%%%%%%%%%%%%

Using two BSS methods, we were able to identify the genuine mid-IR spectra of three 
propulations of carbonaceous nanoparticles in the interstellar medium. We have shown that 
both \emph{FastICA} and NMF are efficient for this task, although NMF is found to be sligthly more appropriate. The extracted spectra enable us to study the evolution of 
carbonaceous nanoparticles in the interstellar medium with unprecedented precision, including 
in external galaxies. These results stress the fact that BSS methods have much to reveal in 
the field of observational astrophysics. We are currently analyzing more spectral cubes
observations from the \emph{Spitzer} database using the strategy presented in this paper.

\bibliography{biblio}

%\begin{thebibliography}{Bibliographie}
\bibliographystyle{splncs}
%\bibitem[1]{rap05}
%Rapacioli, M., C. Jobli, Boissel, P.,\emph{Spectroscopy of polycyclic aromatic hydrocarbons and very small grains in photodissociation regions}, Astronomy and Astrophysics, 429, 193-204, 2005
%\end{thebibliography}

%
\end{document}